\documentclass[11pt,reqno]{amsart}

\usepackage[english]{babel}
\usepackage{amsmath}
\usepackage{amsfonts}
\usepackage{amssymb}
\usepackage{amsthm}
\usepackage{graphicx}
\usepackage{epstopdf}
\usepackage{enumitem}
\usepackage{geometry}
\usepackage{hyperref}
\usepackage{tikz}
\usepackage[all]{xy}
\usepackage{multirow}
\usepackage{anysize}
\usepackage{pgfplots}

\usetikzlibrary{intersections,positioning}

\newtheorem{theorem}{Theorem}[section]
\newtheorem{definition}[theorem]{Definition}

\newcommand{\beq}{\begin{equation}}
\newcommand{\eeq}{\end{equation}}

\marginsize{3.2cm}{3.2cm}{3.2cm}{3.2cm}

\begin{document}

\title{A model of human cooperation in social dilemmas}

\maketitle

\begin{center}VALERIO CAPRARO\\Department of Mathematics\\University of Southampton\\ Southampton, SO17 1BJ, UK\\V.Capraro@soton.ac.uk

\end{center}

\begin{abstract}
Social dilemmas are situations in which collective interests are at odds with private interests: pollution, depletion of natural resources, and intergroup conflicts, are at their core social dilemmas. 

Because of their multidisciplinarity and their importance, social dilemmas have been studied by economists, biologists, psychologists, sociologists, and political scientists. These studies typically explain tendency to cooperation by dividing people in proself and prosocial types, or appealing to forms of external control or, in iterated social dilemmas, to long-term strategies. 

But recent experiments have shown that cooperation is possible even in one-shot social dilemmas without forms of external control and the rate of cooperation typically depends on the payoffs. This makes impossible a predictive division between proself and prosocial people and proves that people have attitude to cooperation by nature.

The key innovation of this article is in fact to postulate that humans have attitude to cooperation by nature and consequently they do not act a priori as single agents, as assumed by standard economic models, but they forecast how a social dilemma would evolve if they formed coalitions and then they act according to their most optimistic forecast. Formalizing this idea we propose the first predictive model of human cooperation able to organize a number of different experimental findings that are not explained by the standard model. We show also that the model makes satisfactorily accurate quantitative predictions of population average behavior in one-shot social dilemmas.\\
\end{abstract}


%\emph{Classification.} Major: Social Sciences; minor: Economic Sciences, Psychological and Cognitive Sciences.\\

Social dilemmas are situations in which collective interests are at odds with private interests ~\cite{1}. In other words, they describe situations in which the fully selfish and rational behavior leads to an outcome smaller than the one the individuals would obtain if they acted collectively. Social dilemmas create then a tension between private interests and public interests, between selfishness and cooperation. Classically, several different social dilemmas have been distinguished, including the Prisoner's dilemma, Chicken, Assurance, Public Goods, the Tragedy of the Commons ~\cite{2}, and, more recently, the Traveler's dilemma ~\cite{3}, ~\cite{4} . Each of these games has been studied by researchers from different disciplines, as economists, biologists, psychologists, sociologists, and political scientists, because of the intrinsic philophical interest in understanding human nature and since many concrete and important situations, as pollution, depletion of natural resources, and intergroup conflict, can be modelled as social dilemmas.\\
The classical approaches explain tendency to cooperation dividing people in proself and prosocial types ~\cite{5}, ~\cite{6}, ~\cite{7}, ~\cite{8}, ~\cite{9}, or appealing to forms of external control ~\cite{10}, ~\cite{11}, ~\cite{12}, or to long-term strategies in iterated social dilemmas~\cite{13}. But, over the years many experiments have been accumulated to show cooperation even in one-shot social dilemmas without external control ~\cite{14}, ~\cite{15}, ~\cite{16}, ~\cite{17}, ~\cite{18}, ~\cite{19}, ~\cite{20}. These and other earlier experiments ~\cite{21}, ~\cite{22}, ~\cite{23}, ~\cite{24} have also shown that the rate of cooperation in the same game depends on the particular payoffs, suggesting that most likely humans are engaged in some sort of indirect reciprocity ~\cite{25}, ~\cite{26} and the same person may behave more or less cooperatively depending on the payoffs. Consequently, the problem of making a predictive division in proself and prosocial types becomes extremely difficult, if not even impossible. \\
From these experiments, we can argue two conclusions: first, the observation of cooperation in one-shot social dilemmas without external controls suggests that the origin of cooperation relies in the human nature; second, the fact that the rate of cooperation depends on the payoffs suggests that it could be computed, at least approximatively, using only the payoffs. The word \emph{approximatively} stands for the fact that numerous experimental studies have shown that cooperation is based on a number of factors, as family history, age, culture, gender, even university course ~\cite{27}, religious beliefs ~\cite{19}, and decision time ~\cite{28}. Therefore, we cannot expect a theory able to say, given only the payoffs, the individual-level rate of cooperation in a social dilemma. We can expect instead a model predicting quite accurately population average behaviour using the mean value of parameters that could be theoretically updated at an individual-level. \\
In this article we make the first step in this direction: (1) we develop the first \emph{predictive} model of cooperation\footnote{We mention that there are many other models that can be applied to explain deviation towards cooperation in social dilemmas, including the cognitive hierarchy model ~\cite{Ca-Ho-Ch04}, the quantal level-k theory ~\cite{St-Wi94}, the level k-theory ~\cite{CG-Cr-Br01}, the quantal response equilibrium ~\cite{MK-Pa95}, the inequity aversion models ~\cite{Fe-Sc99},~\cite{Bo-Oc00} and the noisy introspection model~\cite{Go-Ho04}. Nevertheless, all these models use free parameters and so they are not predictive, but descriptive.}; (2) we show that it explains a number of puzzling experimental findings that are not explained by the standard economic model, such as the fact that the rate of cooperation in the Prisoner's dilemma increases when the cost-benefit ratio decreases, the rate of cooperation in the Traveler's dilemma increases when the bonus/penalty decreases, the rate of cooperation in the Public Goods game increases when the pro-capite marginal return increases, the rate of cooperation in the Chicken game is larger than the rate of cooperation in the Prisoner's dilemma with \emph{similar} payoffs; (3) we show that it makes satisfactorily accurate quantitative predictions of population average behaviour in social dilemmas. 

The key idea behind the model is very simple: since experimental data suggest that humans have attitude to cooperation by nature, we formalize the intuition that people do not act a priori as single agents, but they forecast how the game would be played if they formed coalitions and then they act according to their most optimistic forecast. 

We anticipate that forecasts will be defined by making a comparison between incentive and risk for an agent $i$ to deviate from the collective interest. This comparison leads to associate a probability to the event ``\emph{agent $i$ defects}''. As mentioned, we will show that this procedure works satisfactorily well in the prediction of population average behavior. The problem in passing to individual-level predictions is that the event ``\emph{player $i$ defects}'', given only the payoffs, is not measurable at an individual-level in any universal and objective sense and the dream is to use the factors mentioned above (family history, age, culture, incentives, iterations\footnote{An attempt to extend the present model to iterated social dilemmas has been done in ~\cite{CVPJ13}, leading to very positive results: the predictions get very close to experimental data as the number of iterations increases.}, etc.) to define parameters to update the measure of the event ``\emph{player $i$ defects}'' at an individual-level.

Even though our model is very general and can be applied to every symmetric game, we treat explicitly only four but very relevant and widely studied social dilemmas: the Prisoner's dilemma, the Traveler's dilemma, the Public Goods, and the Tragedy of the Commons. We begin with a short review of these games.

\textbf{Prisoner's dilemma.}  Two players can choose to either ``Cooperate'' or ``Defect''.
If both players cooperate, they both receive the monetary reward, $R$, for cooperating. If one player defects and the other cooperates, then the defector receives the temptation payoff, $T$, while the other receives the sucker payoff, $S$. If both players defect, they both receive the punishment payoff, $P$. Payoffs are subject to the condition $T>R>P>S$.

\textbf{Traveler's dilemma.} Fix a bonus/penalty $b\geq2$. Two travelers have to claim for a reimbursement between $180$ and $300$ monetary units for their (identical) luggage that has been lost by the same air company. The air company wants to avoid that the travelers ask for unreasonably high reimbursements and so it decides to adopt the following rule: the traveler who claims the lowest, say $m$, gets a reimbursement of $m+b$ monetary units, and the other one gets a reimbursement of only $m-b$ monetary units. If both players claim the same, $m$, then they both get reimbursed of $m$ monetary units.

\textbf{Public Goods game.} $N$ agents receive an initial endowment of $y>0$ monetary units and simultaneously choose an amount $0\leq x_i\leq y$ to contribute to a public pool. The total amount in the pot is multiplied by $\alpha_0$ and then divided equally by all group members. So agent $i$ receives a payoff of $u_i(x_1,\ldots,x_N)=y-x_i+\alpha(x_1+\ldots+x_N)$, where $\alpha=\alpha_0/N$. The number $\alpha$ is assumed to belong to the interval $(1/N,1)$ and it is called \emph{constant marginal return}.

\textbf{Tragedy of the commons.} Consider a village with $N$ farmers, that has limited grassland. Each of the $N$ farmers has the option to keep a sheep or not. Let the monetary utility of milk and wool from the sheep be $h>0$. Let the monetary damage to the environment from one sheep grazing over the grassland be denoted by $k_0>0$. Assume $h<k_0<hN$ and let $k=k_0/N$. Let $x_i$ be a variable that takes values $0$ or $1$ and denotes whether the farmer $i$ keeps the sheep or not. The payoff of farmer $i$ is $u_i(x_1,\ldots,x_N)=hx_i  - k (x_1+ \ldots +x_N)$.\\

All these games share the same feature: selfish and rational behavior leads to suboptimal outcomes. In the Prisoner's dilemma, the unique Nash equilibrium is to defect, while both players would be better off if they both cooperate; in the Traveler's dilemma, the unique Nash equilibrium is to claim for the lowest possible amount, producing an outcome smaller than the one they would obtain if they both claim for the largest possible amount; in the Public Goods game, the unique Nash equilibrium is not to contribute anything, while all players would be better off if they all contribute everything; in the Tragedy of the Commons, the unique Nash equilibrium is to keep the sheep, while all farmers would be better off if they all agree not to keep the sheep.

\section{An informal description of the model}\label{se:informal}

Before introducing the model in general, we describe it informally in a particular case. Consider the Prisoner's dilemma (recently experimented using MTurk in ~\cite{20}) with monetary outcomes (expressed in dollars) $T=0.20, R=0.15,P=0.05,S=0$. The idea is that players forecast how the game would be played if they formed coalitions. In a two-player game there are only two possible coalition structures: in the selfish coalition structure $p_s$ players are supposed to follow their private interests and in the cooperative coalition structure $p_c$ they are supposed to follow the collective interest. The analysis of these two coalition structures proceed as follows:
\begin{itemize}
\item In $p_s$ players follow their private interest and therefore, \emph{by definition}, they play the Nash equilibrium $(D,D)$. Since there is no incentive to deviate from a Nash equilibrium, each player gets $0.05$ \emph{for sure} and we say that the value of $p_s$ is $0.05$ and write $v(p_s)=0.05$.
\item To define the value of $p_c$ we argue as follows. If the players follow the collective interest, their largest possible payoff is $0.15$ in correspondence to the profile of strategies $(C,C)$. Since this profile of strategies is not stable (i.e., each player has a non-zero incentive to deviate from it), we introduce a probability to measure how likely such deviations are. To define this probability, we observe that:
\begin{itemize}
\item the \emph{incentive} to deviate from the collective interest is $D(p_c):=0.05$, since each player can get $0.20$ instead of $0.15$, if she defects and the other cooperate; 
\item the \emph{risk} to deviate from the collective interest is $R(p_c):=0.10$, since each player can get only $0.05$ instead of $0.15$ if she follows her private interest but also the other one does the same.
\end{itemize}
We  define the \emph{prior probability} that a player abandons the coalition structure $p_c$ by making a sort of proportion between incentive and risk. Specifically, we define the probability that a player abandons $p_c$ to be $D(p_c)[D(p_c)+R(p_c)]^{-1}$. Now, note that the smallest payoff achievable by a player when she follows $p_c$ but the other player does not is the sucker payoff $S=0$. 
Therefore, we define 
$$
v(p_c)=0.15\cdot\left(1-\frac{D(p_c)}{D(p_c)+R(p_c)}\right)+0\cdot\frac{D(p_c)}{D(p_c)+R(p_c)}=0.10.
$$
\end{itemize}

The numbers $v(p_s)$ and $v(p_c)$ are interpreted as forecasts of the expected payoff for an agent playing according to $p_s$ and $p_c$, respectively. Since $v(p_s)=0.05$ and $v(p_c)=0.10$, the \emph{most optimistic forecast} is in correspondence of the cooperative coalition structure $p_c$.
We use this best forecast to generate common beliefs or, in other words, to make a tacit binding between the players: to play only strategies which give a payoff of at least $0.10$ to both players. More formally, we restrict the set of profiles of strategies and we allow only profiles $\sigma=(\sigma_1,\sigma_2)$, such that $u_i(\sigma)\geq0.10$, for all $i$. We define the cooperative equilibrium to be the unique Nash equilibrium of this restricted game.

\begin{figure}
\begin{centering}
\begin{tikzpicture}
\draw (0,0)node[anchor=north east]{(0,0)}--(3,0)node[anchor=north west]{(1,0)}--(3,3)node[anchor=south west]{(1,1)}--(0,3)node[anchor=south east]{(0,1)}--(0,0);
\draw (1.5,1.5)node[anchor=south east]{};
\draw (2,3)node[anchor=south east]{};
\draw (3,2)node[anchor=south east]{};
\draw (1.5,1.5)node[anchor=north east]{}--(2,3)node[anchor=north west]{};
\draw (1.5,1.5)node[anchor=north east]{}--(3,2)node[anchor=north west]{};
\fill (0,0) circle (.05);
\fill (3,0) circle (.05);
\fill (3,3) circle (.05);
\fill (0,3) circle (.05);
\fill (1.5,1.5) circle (.05);
\fill (3,2) circle (.05);
\fill (2,3) circle (.05);
\fill [red] (1.5,1.5) -- (2,3) -- (3,3)  -- (3,2) -- (1.5,1.5);
\end{tikzpicture}
\end{centering}
\label{restriction}
\caption{The unit square represents the initial set of available profiles of strategies: player 1 can use all strategies $\lambda C+(1-\lambda) D$, for all $\lambda\in[0,1]$; player 2 can use all strategies $\mu C+(1-\mu)D$, for all $\mu\in[0,1]$. The red set represents the set of allowed profiles of strategies in the restricted game.}
\end{figure}
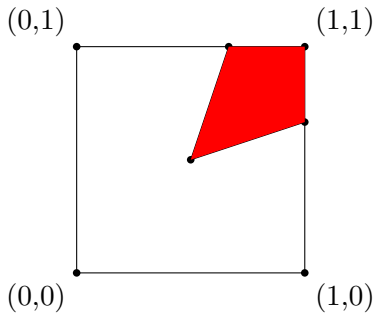

From Fig. \ref{restriction}, it is clear that the cooperative equilibrium is in correspondence of the point in the red set that is closest to $(D,D)$. This point can be computed directly by finding the smallest $\lambda$ such that 

$$
0.15\lambda^2+0.2\lambda(1-\lambda)+0.05(1-\lambda)^2\geq0.1,
$$

that is $\lambda=\frac12$. Consequently, the cooperative equilibrium of this variant of the Prisoner's dilemma is $\frac12 C+\frac12 D$ for both players. Notice that in ~\cite{20} it has been reported that players cooperated with probability 58 per cent in one treatment and 65 per cent in another treatment and the over-cooperation in the second experiment was explained in terms of framing effect due to the different ways in which the same game was presented.

\section{The model}\label{se:model}

We now describe the general model. We recall that, motivated by the observation that attitude to cooperation seems to be intrinsic in the human nature, our main idea is to assume that players do not act a priori as single agents, but they forecast how the game would be played if they formed coalitions and then they play according to their most optimistic forecast. The only technical difficulty to formalize this idea is to define the \emph{forecasts}. Following the example described in the previous section, they will be defined by assigning to each player $i$ and to each partition $p$ of the player set $P$, interpreted as a possible coalition structure, a number $v_i(p)$ which represents the expected payoff of player $i$ when she plays according to the coalition structure $p$. This value will be indeed defined as an average 
$$
v_i(p)=\sum_{J\subseteq P\setminus\{i\}}e_{i,J}(p)\tau_{i,J}(p),
$$
where $\tau_{i,J}(p)$ represents the \emph{prior probability} that players $i$ assigns to the event ``\emph{players in $J$ abandon the coalition structure $p$}'' and $e_{i,J}(p)$ is the infimum of payoffs of player $i$ when she plays according to the coalition structure $p$ and players in $J$ abandon the coalition.

This idea is very general and indeed, in a long-term working paper, we are developping the theory for every normal form game ~\cite{29}. In case of the classical social dilemmas in consideration the theory is much easier, because of their symmetry.
\begin{itemize}
\item \textbf{Symmetry.} All players have the same set of strategies $S$ and for each player $i$, for each permutation $\pi$ of the set of players and for each $(\sigma_1,\ldots,\sigma_N)\in S^N$ one has 
\begin{align}
u_i(\sigma_1,\ldots,\sigma_N)=u_{\pi(i)}(\sigma_{\pi(1)},\ldots,\sigma_{\pi(N)}). 
\end{align}
\end{itemize}

Coming to the description of the model, let $\mathcal G$ be a symmetric game and denote $P$ the set of players, each of which has pure strategy set $S_i$, mixed strategies $\mathcal P(S_i)$ and payoff function $u_i$. We start by assuming, for simplicity, that $P=\{1,2\}$ and we will explain, at the end of this section, how the model generalizes to $N$-player games.

A \emph{coalition structure} is a partition $p$ of the set of players, that is a collection of pairwise disjoint subsets of $P$ whose union covers $P$ . Every set in the partition is called \emph{coalition}. Given a coalition structure $p$, we denote by $\mathcal G_p$ the game associated to $p$, whose players in the same coalition play as a single player whose payoff is the sum of the payoffs of the players belonging to that coalition. Call $\text{Nash}(\mathcal G_p)$ the set of Nash equilibria of the game $\mathcal G_p$. Now fix $i\in P$ and let $-i$ denote the other player. We denote by $D_{-i}(p)$ the maximal payoff that player $-i$ can obtain leaving the coalition structure $p$. Formally, 
\begin{align}\label{eq:incentive}
D_{-i}(p):=\sup\{u_{-i}(\sigma_i^p,\sigma_{-i})-u_{-i}(\sigma_i^p,\sigma_{-i}^p) : \sigma_{-i}\in\mathcal P(S_{-i}), \sigma^p\in\text{Nash}(\mathcal G_p)\}.
\end{align}

 $D_{-i}(p)$ will be called \emph{incentive} of player $-i$ to abandon the coalition structure $p$.

Given a profile of strategies $(\sigma_1,\sigma_2)$, a strategy $\sigma_i'\in\mathcal P(S_i)$ is called $i$-deviation from $(\sigma_1,\sigma_2)$ if $u_i(\sigma_{i}',\sigma_{-i})\geq u_i(\sigma_1,\sigma_2)$.  

We denote by $R_{-i}(p)$ the maximal loss that players $-i$ can incur if she decides to leave the coalition structure $p$ to try to achieve her maximal possible gain, but also player $i$ deviates from the coalition structure $p$ either to follow her selfish interests or to anticipate player $-i$'s deviation. Formally, 
\begin{align}\label{eq:risk}
R_{-i}(p):=\sup\{u_{-i}(\sigma_i^p,\sigma_{-i}^p)-u_{-i}(\sigma_i,\sigma_{-i})\},
\end{align}

 where $\sigma^p$ runs over the set of Nash equilibria of $\mathcal G_p$ and, for each such $\sigma^p$, $\sigma_{-i}$ runs over the set of strategies such that $u_{-i}(\sigma_i^p,\sigma_{-i})$ is maximized and $\sigma_i$ runs over the strategies that are $i$-deviations from either $(\sigma_i^p,\sigma_{-i}^p)$ or $(\sigma^p_i,\sigma_{-i})$. $R_{-i}(p)$ is called \emph{risk} for player $-i$ in abandoning the coalition structure $p$.

We define the probability of deviating from the coalition structure $p$ by making a comparison between incentive and risk. There are certainly many ways to do such comparison. In this paper we use a quite intuitive and seemingly natural way to make it and, in future research, it would be important to investigate some others. Specifically, we define

\begin{align}\label{eq:probability}
\tau_{i,-i}(p):=\frac{D_{-i}(p)}{D_{-i}(p)+R_{-i}(p)}
\end{align}

and we interpret this number as \emph{prior} probability that player $i$ assigns to the event ``\emph{player $-i$ abandons the coalition structure $p$}''. Therefore $\tau_{i,\emptyset}(p):=1-\tau_{i,-i}(p)$ is interpreted as \emph{prior} probability that nobody abandons the coalition structure $p$. Now, let $e_{i,\emptyset}(p)$ be the infimum of payoffs for player $i$ if nobody abandons the coalition structure $p$, that is the infimum of payoffs for player $i$ when each player plays according to a Nash equilibrium of $\mathcal G_p$, and let $e_{i,-i}(p)$ be the infimum of payoffs of player $i$ when she plays according to a Nash equilibrium of $\mathcal G_p$ and $-i$ plays a $(-i)$-deviation from a Nash equilibrium of $\mathcal G_p$. The value for player $i$ of the coalition structure $p$ is by definition 
\begin{align}
v_i(p):=e_{i,\emptyset}(p)\tau_{i,\emptyset}(p)+e_{i,-i}(p)\tau_{i,-i}(p).
\end{align} 
Symmetry implies that $v_i(p)=v_j(p)=:v(p)$, for all $i,j\in P$. Consequently, there is a coalition structure $\overline p$ (independent of $i$) which maximizes $v(p)$. We use the number $v(\overline p)$ to define common beliefs or, in other words, to make a tacit binding among the players. 

\begin{definition}
{\rm The induced game $\text{Ind}(\mathcal G,\overline p)$ is the same game as $\mathcal G$ except for the set of allowed profiles of strategies: in the induced game only profiles of strategies $\sigma=(\sigma_1,\sigma_2)$ such that $u_i(\sigma)\geq v(\overline p)$, for all $i$, are allowed. }
\end{definition}
Observe that the induced game does not depend on the maximizing coalition structure, that is, in case of multiple coalition structures maximizing the value, one can choose one of them casually to define the induced game and this game does not depend on such choice.

Since the set of allowed strategies in the induced game is convex and compact (and non-empty) one can compute Nash equilibria of the induced game.
\begin{definition}
{\rm A cooperative equilibrium for $\mathcal G$ is a Nash equilibrium of the game $\text{Ind}(\mathcal G,\overline p)$.}
\end{definition}

Observe that this model implicitly assumes that it is common knowledge that both players apply the same method of reasoning, that is, each player knows that the other player thinks about coalitions when making her decision. As we elaborate in Section \ref{se:conclusions}, we believe that this assumption is not unreasonable and may provide a realistic picture of the mental processes that real subjects perform during the game.

In case of $N$-player games the idea is to define $\tau_{i,j}(p)$ for every single player $j\neq i$ and then use the law of total probabilities to extend this measure to a probability measure on the set $P\setminus\{i\}$. To use the law of total probabilities we need to know the probabilities that two or more given players deviate from $p$. This is easy in situations of perfect anonimity: one can just assume that the events ``player $j$ deviates'' and ``player $k$ deviates'' are independent and then multiply the respective probabilities. The situation where a player may influence the choice of another player is much more interesting and worthy of being explored.

Finally, we observe that the N-person classical social dilemmas in consideration are computationally very simple, since it is enough to study only the fully selfish coalition structure $p_s$ (in which all players play according to a Nash equilibrium of the original game) and the fully cooperative coalition structure $p_c$ (in which all players play collectively). More formally, given a coalition structure $p\neq p_s,p_c$, one has $v(p)\leq v(p_c)$. Therefore, in order to find a coalition structure that maximizes the value, it is enough to know the values $v(p_s)$ and $v(p_c)$.
\section{Predictions of the model}\label{se:predictions}

\textbf{Prisoner's dilemma.} We compute the cooperative equilibrium of the Prisoner's dilemma  in two variants, starting from the one already discussed in Section \ref{se:informal} with monetary outcomes (expressed in dollars) $T=0.20, R=0.15,P=0.05,S=0$.
%Fix $i=1$. Denote by $p_s$ the selfish coalition structure, where the players are supposed to act separately. Then $\mathcal G_{p_s}=\mathcal G$, whose unique Nash equilibrium is $(D,D)$. Since a Nash equilibrium has no deviations, then $D_2(p_c)=0$ and consequently $v(p_s)=0.05$. Now, let $p_c$ be the cooperative coalition structure, where the players are supposed to play together. The game $\mathcal G_{p_c}$ is a one-player game whose only Nash equilibrium is $(C,C)$. Now, $D_2(p_c)=0.05$, since the second player can get $0.20$ instead of $0.15$ if she defects and the first player cooperates, and $R_2(p_c)=0.10$, since the second player risks to get $0.05$ instead of $0.15$ if also the other player defects. Finally $e_{1,\emptyset}(p_c)=0.15$ and $e_{1,2}(p_c)=0$. Consequently, $v(p_c)=0.10$, that is larger than $v(p_s)$. So we need to compute the Nash equilibrium of $\text{Ind}(\mathcal G,p_c)$. By symmetry of the game, this is the same as finding the smallest $\lambda$ such that $0.15\lambda^2+0.2\lambda(1-\lambda)+0.05(1-\lambda)^2\geq0.1$, that is $\lambda=\frac12$. 
In this case, the reader can easily check, following the computation sketched in Section \ref{se:informal}, that the cooperative equilibrium is $\frac12 C+\frac12 D$ for both players. Notice that in ~\cite{20} it has been reported that players cooperated with probability 58 per cent in one treatment and 65 per cent in another treatment and the over-cooperation in the second experiment was explained in terms of framing effect due to the different ways in which the same game were presented.

Similar results can be obtained making a comparison between the experimental data reported in ~\cite{19} on the one-shot prisoner's dilemma with $T=10, R=7,P=3, S=0$ and its cooperative equilibrium: 37 per cent of subjects cooperated in the laboratory, while the cooperative equilibrium is $\frac14 C+\frac34 D$. We mention that the same experiment was repeated using MTurk and ten times smaller outcomes, giving a slightly larger percentage of cooperation (47 per cent). Nevertheless, it was shown in ~\cite{19} that this difference was not statistically significant.

Now we consider a parametric Prisoner's dilemma. Fix $k>0$ and consider the following monetary outcomes: $T=k+2, R=k+1, P=1, S=0$. The intuition suggests that people should be perfectly selfish for $k=0$, they should get more cooperative as $k$ increases and they should tend to be perfectly cooperative as $k$ approaches infinity. This qualitative behavior was indeed observed in iterated treatments in ~\cite{18}.

We show that this is in fact the behavior of the cooperative equilibrium. Indeed, one obtains that the cooperative equilibrium coincides with Nash equilibrium for $k\leq1$, while, for $k>1$, it is
$$
\frac{k-1}{k}C+\frac{1}{k}D,
$$
which moves continuously and monotonically from defection to cooperation as $k$ increases and tends to cooperation as $k$ tends to infinity. Note that the fact that the cooperative equilibrium coincides with Nash equilibrium for $k\leq1$ shows also that Nash equilibrium and cooperative equilibrium are not disjoint solution concepts. Colloquially speaking, players get selfish when they understand that cooperating is not fruitful.

\textbf{Traveler's dilemma.} One has $v(p_s)=180$, since $(180,180)$ is the unique Nash equilibrium of the Traveler's dilemma, and 
$$
v(p_c)=300\cdot\frac{b+2}{2b+1}+(300-2b)\cdot\frac{b-1}{2b+1}.
$$
since the unique Nash equilibrium of $\mathcal G_{p_c}$ is $(300,300)$, $D_2(p_c)=b-1$, in correspondence of $(300,299)$, $R_2(p_c)=b+2$ in correspondence of $(298,299)$, $e_{1,2}(p_c)=300-2b$ in correspondence of $(300,300-b)$, and clearly $e_{1,\emptyset}(p_c)=300$ in correspondence of $(300,300)$.

Consequently the cooperative equilibrium strongly depends on $b$: the predicted claims get smaller as $b$ get larger. In other words, cooperation is more difficult as the bonus/penalty increases. This behaviour has been indeed qualitatively observed both in one-shot and iterated games ~\cite{30},~\cite{16}, ~\cite{17}, and ~\cite{31}. We are aware of only two experimental studies devoted to one-shot Traveler's dilemmas. For these experiments, the prediction of the the cooperative equilibrium are even quantitatively close. Indeed, (1) for $b=5$ one finds that the unique cooperative equilibrium is a suitable convex combinations of the strategies 296 and 297. This meets the experimental data reported in ~\cite{16}, where they observed that about 80 per cent of the subjects played a strategy between 290 and 300 with an average of 295; (2) For $b=180$, one has $v(p_c)<v(p_s)$, and then the cooperative equilibrium coincides with the Nash equilibrium. This matches the experimental data reported in ~\cite{16}, where they observed that about 80 per cent of the players played the Nash equilibrium; (3) For $b=2$ and strategy sets $\{2,3,\ldots,100\}$, in ~\cite{17} it has been reported that 38 out of 45 game theorists chose a strategy between 90 and 100 and 28 of them chose a strategy between 97 and 100. In this case $v(p_c)=99.2$ and therefore the cooperative equilibrium is close to the pure strategy $99$.

\textbf{Public good game.} The unique Nash equilibrium is $x_i=0$, for all $i$, in correspondence of which each player gets $y$. Consequently $v(p_s)=y$. On the other hand, one has
$$
v(p_c)=2\alpha y\cdot\frac{2\alpha -1}{\alpha }+\alpha y\cdot\frac{1-\alpha}{\alpha}=(3\alpha-1)y.
$$ 
Therefore, $v(p_c)\leq v(p_s)$ if and only if $\alpha\leq\frac23$. In other words, when $\alpha$ is small - recall that $\alpha$ is assumed to belong to the interval $(\frac{1}{2},1)$ - the cooperative equilibrium reduces to Nash equilibrium and the larger is $\alpha$ the larger is the rate of cooperation predicted by the cooperative equilibrium. The fact that human behavior depends on $\alpha$ in this way has been indeed observed several times (see, e.g., ~\cite{14}, \cite{32}). As a quantitative comparison, we consider the experimental data reported in ~\cite{33}, with $\alpha=0.8$. We normalize $y$ to be equal to 1 (in the experiment $y=0.04$ dollars). In this case the cooperative equilibrium is \emph{supported between} 0.66 and 0.67. In ~\cite{33} it has been reported that the average of contributions was 0.50, but the mode was 0.60 (6 out of 32 times) followed by 0.80 (5 out of 32 times). 

\textbf{Tragedy of the Commons.} One easily sees that the Tragedy of the Commons and the Public Goods game represent the same strategic situation, just by setting $\alpha:=\frac{k}{h}$, that can be interpreted as the effective cost of having a sheep. In particular, one finds that $v(p_c)>v(p_s)$ if and only if $\alpha>\frac23$.

\textbf{Comparison between the Prisoner's dilemma and Chicken.} It has been observed in ~\cite{34} that the rate of cooperation in the iterated Prisoner's dilemma is significantly less than the rate of cooperation in the iterated Chicken game\footnote{We recall that the Chicken game is basically the same as the Prisoner's dilemma except for the fact that payoffs are subject to the condition $T>R>S>P$. The Chicken game has two pure Nash equilibria, $(C,D)$ and $(D,C)$, and a symmetric evolutionarily stable mixed Nash equilibrium depending on the payoffs. Observe that $e_{i,\emptyset}(p_s)=P$, since it is the infimum of payoffs of player $i$ when each player plays in according to a Nash equilibrium. Such infimum is attained in correspondence to the profile of strategies $(D,D)$.} with \emph{similar} payoffs, that is, with payoffs such that the average payoffs across outcomes is the same in both games. We now show that this behavior is predicted by the cooperative equilibrium in one-shot games, giving a \emph{qualitative explanation} of why we observe more cooperation in the iterated Chicken game than in the iterated Prisoner's dilemma \footnote{The expression \emph{qualitative explanation} stands for the fact that, of course, a direct comparison between iterated and one-shot games cannot be done, since the former have a much richer set of strategies. Nevertheless, we find quite remarkable the fact that this difference in behavior observed in iterated treatments is predicted for one-shot treatments: we believe that this connection is not casual and deserves to be investigate better.}. The payoffs used in ~\cite{34} are $T=400, R=300, D=0, S=-100$ for the Prisoner's dilemma and  $T=300, R=200, S=100, D=0$ for the Chicken game. One finds that the cooperative equilibrium of this variant of the Prisoner's dilemma is $\frac23 C+\frac13 D$ and the cooperative equilibrium of this variant of the Chicken game coincides with the evolutionarily stable strategy $\frac67 C+\frac17 D$. So the rate of cooperation predicted by the cooperative equilibrium is significantly higher in the Chicken game.

\section{Conclusions}\label{se:conclusions}
Many experiments over the years have shown that humans may act cooperatively even in one-shot social dilemmas without forms of external controls and the rate of cooperation depends on the payoffs. This suggests that humans have attitude to cooperation by nature and therefore they do not act a priori as single players, as typically assumed in economics, but they forecast how the game would be played if they formed coalitions and then they play according to their most optimistic forecast. 

We have formalized this idea assuming that each player makes an evaluation of the probability that another player abandons the collective interest to follow her private interest. This probability is defined by making a comparison between incentive and risk to deviate from the collective interest and gives rise to common beliefs that, mathematically, correspond to define a suitable restriction of the original game. On the one hand, this procedure seems \emph{qualitatively} reasonable and we believe it provides a realistic picture of the mental processes that real subjects perform during the game. On the other hand, the formalization of this process, that is, the definitions of the risk, incentive, probabilities, and the induced game, is mathematically simple and seemingly natural but certainly deserves to be investigated better and possibly improved in future research.

 However, the actual model makes us optimistic about this direction of research, being the first predictive model able to: (1) make satisfactorily accurate predictions of population average behavior in social dilemmas; (2) explain a number of experimental findings, such as the fact that the rate of cooperation in the Prisoner's dilemma increases when the cost-benefit ratio decreases, the rate of cooperation in the Traveler's dilemma increases when the bonus/penalty decreases, the rate of cooperation in the Public Goods game increases when the pro-capite marginal return increases, the rate of cooperation in the Chicken game is larger than the rate of cooperation in the Prisoner's dilemma with \emph{similar} payoffs. 

The dream is to incorporate other components (as family history, age, culture, incentives, iterations, etc.) into the model in order to make individual-level predictions.


\begin{thebibliography}{9}
\bibitem{1} Kerr NL (1983) \emph{Motivation losses in small groups: A social dilemma analysis}, Journal of Personality and Social Psychology 45:819-828.
\bibitem{2} Kollock P (1988) \emph{Social dilemmas: Anatomy of cooperation}, Annual Review of Sociology 24:183-214.
\bibitem{3} Basu K (1994) \emph{The Traveler's Dilemma: Paradoxes of Rationality in Game Theory}, American Economic Review 84 (2): 391-395.
\bibitem{4} Manapat ML, Rand DG, Pawlowitsch C, Nowak MA (2012) \emph{Stochastic evolutionary dynamics resolve the Traveler’s Dilemma}. Journal of Theoretical Biology 303:119-127.
\bibitem{5} Liebrand WBG (1984) \emph{The effect of social motives, communication and group size on behavior in an n-person multi-stage mixed-motive game}, Eur. J. Soc. Psychol. 14:239-264.
\bibitem{6} Liebrand WBG, Wilke HAM, Vogel R, Wolters FJM  (1986) \emph{Value orientation and conformity in three types of social dilemma games}, J. Conflict Resolut. 30:77-97.
\bibitem{7} Kramer RM, McClintock CG, Messick DM (1986) \emph{Social values and cooperative response to a simulated resource conservation crisis}, J. Pers. 54:576-591.
\bibitem{8} Kuhlman DM, Camac CR, Cunha DA (1986) \emph{Individual differences in social orientation}, In Experimental Social Dilemmas, ed. HAM Wilke, DM Messick, C Rutte, pp. 151-174. Frankfurt: Verlag Peter Lang.
\bibitem{9} McClintock CG, Liebrand WBG (1988) \emph{Role of interdependence structure, individual value orientation, and another’s strategy in social decision making: a transformational analysi},. J. Pers. Soc. Psychol. 55 (3):396-409.
\bibitem{10} Olson M (1965) \emph{The Logic of Collective Action: Public Goods and the Theory of Groups}, Cambridge, MA: Harvard Univ. Press.
\bibitem{11} Hardin G (1968) \emph{The tragedy of the commons}, Science 162: 1243-1248.
\bibitem{12} Dawes R (1980) \emph{Social dilemmas}, Annu. Rev. Psychol. 31:169-193.
\bibitem{13} Axelrod R (1984) \emph{The Evolution of Cooperation}, New York: Basic Books.
\bibitem{14} Isaac MR, Walker J (1988) \emph{Group size effects in public goods provision: The voluntary contribution mechanism}, Quarterly Journal of Economics 103:179-200.
\bibitem{15} Cooper R, DeJong DV, Forsythe R, Ross TW (1996) \emph{Cooperation without Reputation: Experimental Evidence from Prisoner’s Dilemma Games}, Games and Economic Behavior 12:187-218.
\bibitem{16} Goeree J, Holt C (2001) \emph{Ten Little Treasures of Game Theory and Ten Intuitive Contradictions}, American Economic Review 91:1402-1422.
\bibitem{17} Becker T, Carter M, Naeve J (2006) \emph{Experts Playing the Travelers Dilemma}, Working Paper 252, Institute for Economics, Hohenheim University.
\bibitem{18} Dreber A, Rand DG, Fudenberg D, Nowak MA (2008) \emph{Winners don't punish}, Nature 452:348-351.
\bibitem{19} Horton JJ, Rand DG, Zeckhauser RJ (2011) \emph{The online laboratory: conducting experiments in a real labor market}, Experimental Economics 14:399-425.
\bibitem{20} Dreber A, Ellingsen T, Johannesson M, Rand DG (2012) \emph{Do People Care About Social Context? Framing Effects in Dictator Games}. Experimental Economics doi:10.1007/s10683-012-9341-9. 
\bibitem{21} Kelley HH, Grzelak J (1972) \emph{Conflict between individual and common interest in an N-person relationship}, J. Pers. Soc. Psychol. 21:190-197.
\bibitem{22} Bonacich P, Shure G, Kahan J, Meeker R (1976) \emph{Cooperation and group size in the n- person prisoner’s dilemma}, J. Conflict Resolution 20:687-706.
\bibitem{23} Komorita SS, Sweeney J, Kravitz DA (1980) \emph{Cooperative choice in the n-person dilemma situation}, J. Pers. Soc. Psychol. 38:504-516.
\bibitem{24} Isaac RM, Walker J, Thomas S. (1984) \emph{Divergent evidence on free riding: an experimental examination of possible explanations}, Public Choice 43:113-149.
\bibitem{25} Nowak MA, Sigmund K (1998) \emph{Evolution of indirect reciprocity by image scoring}, Nature 393:573-577.
\bibitem{26} Nowak MA (2006) \emph{Five rules for the evolution of cooperation}, Science 314:1560-1563.
\bibitem{27} Marwell G, Ames RE (1981) \emph{Economists free ride, does anyone else?}, Journal of Public Economics 15:295-310.
\bibitem{28} Rand DG, Greene JD, Nowak MA (2012) \emph{Spontaneous giving and calculated greed}. Nature 489:427-430.
\bibitem{Ca-Ho-Ch04} Camerer, C., Ho, T., Chong, J. {A cognitive hierarchy model of games}.  Quaterly J. of Economics 119 (3), 861--898 (2004)
\bibitem{MK-Pa95} McKelvey, R., Palfrey, T. {Quantal response equilibria for normal form games}. Games and Economic Behavior 10 (1), 6--38 (1995)
\bibitem{St-Wi94} Stahl, D., Wilson, P. {Experimental evidence on players' models of other players}.  J. Economic Behavior and Organization 25 (3), 309--327 (1994)
\bibitem{CG-Cr-Br01} Costa-Gomes, M., Crawford, V., Broseta, B. {Cognition and behavior in normal form games: An experimental study}. Econometrica 69 (5), 1193--1235 (2001)
\bibitem{Fe-Sc99} Fehr, E. and K. Schmidt, \emph{A theory of fairness, competition and cooperation}, Quaterly Journal of Economics 114 (3) (1999), 817-868.
\bibitem{Bo-Oc00} Bolton, G.E. and A. Ockenfels, \emph{ERC: A Theory of Equity, Reciprocity and Competition}, The American Economic Review 90 (2000) 166-193.
\bibitem{Go-Ho04} Goeree JK, Holt, CA (2004) \emph{A model of noisy introspection}. Games and Economic Behavior 46:365-382.
\bibitem{CVPJ13} Capraro V, Venanzi M, Polukarov M, Jennings NR (2013) \emph{Cooperative equilibria in iterated social dilemmas.} Submitted preprint available at SSRN: http://ssrn.com/abstract=2268998.
\bibitem{29} Capraro V (2013) \emph{A solution concept for games with altruism and cooperation}, Working Paper available at http://arxiv.org/pdf/1302.3988.pdf. 
\bibitem{30} Capra M, Goeree JK, Gomez R, Holt CA (1999) \emph{Anomalous Behavior in a Traveler’s Dilemma?}, American Economic Review, Vol. 89 (3):678-690.
\bibitem{31} Basu K, Becchetti L, Stanca L  (2011) \emph{Experiments with the Traveler’s Dilemma: welfare, strategic choice and implicit collusion}, Social Choice and Welfare 37 (4):575-595.
\bibitem{32} Gunnthorsdottir A, Houser D, McCabe K (2007) \emph{Disposition, history and contributions in public goods experiments}, Journal of Economic Behavior and Organization 62:304-315.
\bibitem{33} Goeree JK, Holt CA, Laury SK (2002) \emph{Private Costs and Public Benefits: Unraveling the Effects of Altruism and Noisy Behavior}, Journal of Public Economics, 83 (2):255-276.
\bibitem{34} K\"ummerli R, Colliard C, Fiechter N, Petitpierre B, Russier F, Keller L (2007) \emph{Human cooperation in social dilemmas: comparing the Snowdrift game with the Prisoner's Dilemma}, Proc. R. Soc. B 274 (1628):2965-2970.


 

\end{thebibliography}
\end{document}